\documentclass[a4paper]{jpconf}
\usepackage{graphicx}
\begin{document}
\title{Neutrino astrophysics and its connections to nuclear physics}

\author{Maria Cristina Volpe}

\address{Astro-Particule
et Cosmologie (APC),
CNRS UMR 7164, Universit\'e Denis Diderot,\\
10, rue Alice Domon et L\'eonie Duquet,
75205 Paris Cedex 13, France}

\ead{volpe@apc.in2p3.fr}

\begin{abstract}
We highlight recent developments in neutrino astrophysics. We  discuss some of the connections with nuclear physics. 
\end{abstract}

\section{Milestones in neutrino astrophysics}
Neutrino astronomy started with the pioneering measurement of solar neutrinos by R. Davis' experiment in the sixties \cite{Davis:1968cp} and the observation of neutrinos from the supernova 1987A located in the Large Magellanic Cloud by Kamiokande II \cite{Hirata:1987hu}, by IMB \cite{Bionta:1987qt} and Baksan \cite{Alekseev:1988gp}. Evidence of ultra-high energy neutrinos in the PeV energy range by the ICECUBE Collaboration has initiated high energy neutrino astronomy \cite{Aartsen:2014gkd}.
Moreover, the measurements of gravitational waves from binary black holes \cite{Abbott:2016blz}
and neutron star mergers \cite{TheLIGOScientific:2017qsa} has opened the era of multi-messenger astronomy.

Neutrinos are elementary massive particles with mixings. 
This is known since the discovery of vacuum oscillation by the Super-Kamiokande experiment \cite{Fukuda:1998mi}, that had an impact in particle physics, astrophysics and cosmology.  
This phenomenon occurs because the mass or propagation basis and the flavor or interaction basis do not coincide. They are related by the Pontecorvo-Maki-Nakagawa-Sakata unitary matrix \cite{Maki:1962mu} which depends on three
mixing angles, one Dirac and two Majorana phases, if there exists only three active neutrino flavors.
If neutrinos are Dirac particles, the neutrino mass can be implemented by adding a right-handed singlet to the Glashow-Weinberg-Salam model. If neutrinos are Majorana particles, significant extensions to the standard model are necessary that produce total lepton number violation (see e.g. the book \cite{Giunti:2007ry}).

First identified by R. Davis' experiment, the solar neutrino deficit problem has been solved with the neutrino oscillation discovery \cite{Fukuda:1998mi}, the SNO measurement of the total solar neutrino flux in agreement with expectations from the Standard Solar Model \cite{Ahmad:2002jz} and with the identification of the Large-Mixing-Angle (LMA) solution by KamLAND \cite{Eguchi:2002dm} (see e.g. \cite{Robertson:2012ib} for a review). In particular, the production of solar neutrinos and their flavor conversion are well understood. 
Historically, the study of solar $\nu$ has played a major role in establishing how energy is produced in low mass main sequence stars, as our Sun, and in ascertaining  neutrino flavor conversion. The precise measurement of neutrinos coming from the Sun's interior have confirmed that the proton-proton (pp) reaction chain, that burns hydrogen to produce helium-4 nuclei, is responsible for $99 \%$ of energy production in the Sun \cite{Bellini:2014uqa}. The (hopefully) future measurement of neutrinos from the carbon-nitrogen-oxygen (CNO) cycle would provide us with the first direct observation of the main energy production mechanism in high mass main sequence stars \cite{Bethe:1939bt,Robertson:2012ib,Villante:2014txa}.

Concerning flavor conversion in astrophysical environments, the observation of a reduced electron neutrino flux by solar experiments and of the total solar flux by the SNO experiment has demonstrated that : {\it i)} solar neutrinos with less than 2 MeV energy undergo averaged vacuum oscillations; {\it ii)} electron neutrinos than have more than 2 MeV energy convert into muon and tau neutrinos because of the Mikheev-Smirnov-Wolfenstein (MSW) effect \cite{Wolfenstein:1977ue,Mikheev:1986gs,Bellini:2014uqa}. This is due to neutrino-matter interactions that neutrinos encounter while propagating through the medium. They can be accounted for, by using the mean-field approximation. Due to this mean field,
$\nu_e$ efficiently (adiabatically) 
change into $\nu_{\mu}, \nu_{\tau}$ at the MSW resonant location that depends on the solar electron number density profile, the neutrino energy and mixing parameters. The MSW effect is analogous to the two-level problem in quantum mechanics. 

Since 1998 experiments exploiting athmospheric, solar, reactor and accelerator neutrinos 
have allowed to determine the three neutrino mixing angles and two mass-squared differences responsible for vacuum oscillations. 
Still, some anomalies need to be elucidated, that do not fit into the theoretical framework of three active neutrino flavors, such as the so-called "reactor anomaly". They could be explained by the existence of sterile neutrinos that several experiments are searching for ( see e.g. \cite{Giunti:2015wnd}).
Key open questions on neutrino properties remain to be answered by upcoming experiments, including the absolute mass scale, the neutrino mass ordering, the neutrino nature  (Dirac versus Majorana), the existence of sterile neutrinos and of CP violating phases. The combined analysis of available experimental data appear to favor normal (versus inverted)) mass ordering\footnote{The mass ordering is unknown since one of the mass-squared differences has not been measured yet (referred to as the atmospheric one). As a consequence we do not yet know if the third mass eigenstate is the lightest or the heaviest.} and a CP violating Dirac phase around 1.5 $\pi$ \cite{Capozzi:2017ipn}. 
 
Neutrinos are copiously produced in stellar environments. Weakly interacting, they propagate from stellar cores, violent phenomena such core-collapse supernovae, or in accretion disks around compact binaries (i.e. black hole-black hole, neutron star-black hole, neutron star-neutron star). Their fluxes and spectra encode information on these environments and on flavor conversion in such media.  In the last decade, theoretical studies have shown conversion phenomena that occur due to the neutrino-neutrino interaction, the presence of shock waves, of turbulence, or of new physics, like non-standard neutrino-matter or neutrino-neutrino interactions. Pinning down the underlying mechanisms and the conditions for their occurrence requires demanding simulations, much more complex than in the solar case.
Such phenomena can potentially impact the explosion dynamics of iron core-collapse supernovae and the conditions for the nucleosynthesis of heavy elements ($r$-process) as well as induce neutrino spectral 
swapping(s) in future observations both of (extra)galactic supernova neutrinos and of the diffuse supernova neutrino background. The former are rare events, typically one every thirty years, while the latter should be discovered by the EGADS project \cite{Xu:2016cfv,Beacom:2003nk}.

\section{Neutrino flavor evolution as a many-body problem}
A gas of self-interacting neutrinos and antineutrinos propagating in a dense astrophysical environment is a many-body system. Through many-body techniques formal connections can be established between such as neutrino gas and other many-body systems like atomic nuclei, or condensed matter. In particular the dynamics of the gas can be treated using the well known Born-Bogoliubov-Green-Kirkwood-Yvon (BBGKY) hierarchy. 

BBGKY is a system of coupled first-order integro-differential equations for reduced density matrices, or equal-time correlation functions. Originally introduced
for non-relativistic systems, the BBGKY hierarchy generalisation to relativistic systems involves an infinite set of equations. The non-relativistic version of the hierarchy has been used to describe atomic nuclei, their dynamics e.g. in nucleus-nucleus collisions, and collective excitations, i.e. giant resonances \cite{lacroix}.
  
The first of BBGKY equations determines the evolution of the one-body (or single-particle) density matrix leading to the time-dependent Hartree-Fock equation (TDHF), when one neglects the correlated part of the two-body density matrix \cite{lacroix}. Its inclusion brings to the Boltzmann equation that includes collisions in the dynamics. The linearised version of the TDHF equation is the Random-Phase-Approximation (RPA) used for the description of collective excitations such as the giant dipole resonance where protons and neutrons oscillate against each other. 
This collective vibration is described within RPA which is an eigenvalue equation for the particle-hole and hole-particle excitations accounted for by the forward $X$ and backward $Y$ amplitudes  \cite{lacroix}. Its eigenenergies describe the frequencies of such modes. However RPA can have complex eigenvalues when the starting point (the nuclear ground state) is not the lowest energy state. Such complex values correspond to instabilities of the system.

Ref.\cite{Volpe:2013uxl} has applied the relativistic version of BBGKY to the description of a gas of neutrinos and antineutrinos evolving in an astrophysical environment. The authors have shown that the first equation gives the Liouville Von Neumann equation for single-particle densities that is commonly employed to study neutrino flavor evolution in media (see e.g. \cite{Serreau:2014cfa}). 
In this case, BBGKY has to be applied to reduced density matrices since off-diagonal contributions have to be introduced to account for the neutrino mixings.
The mean-field associated with the particles composing the medium can be shown to give various terms. For example the well known MSW potential $V_{CC} = \sqrt{2} G_F \rho_e (r)$\footnote{The quantity $\rho_e$ is the electron number density that depends on the medium.} arises because of the charged-current neutrino interaction with electrons. Similarly one obtains the non-linear neutrino self-interaction potential from neutrino interaction with the neutrinos composing the medium. Such a potential has off-diagonal contributions, as first pointed out by Pantaleone \cite{Pantaleone:1992eq}, because of the neutrino mixings.

Ref.\cite{Vaananen:2013qja} has provided a linearised version of the neutrino evolution equations  in astrophysical media and shown that it is as RPA used for atomic nuclei, when one replaces the $X$ and $Y$ amplitudes with the off-diagonal terms of the neutrino and antineutrino density matrix respectively. Another difference between the nuclear case and the neutrino case is that in the latter case the starting point are the neutrino fluxes at the neutrinosphere that are
quasi-stationary solutions of the evolution equation while in the former the nuclear ground state is a stationary solution. Moreover for neutrinos one seeks mostly for unstable solutions that correspond to flavor modification whereas for nuclei one looks for stable solutions corresponding to nuclear excitations.

The investigation of neutrino self-interactions has triggered a decade of studies aiming at understanding a plethora of novel flavor conversion phenomena due to its presence \cite{Duan:2010bg}.
In fact, when a core-collapse supernova explodes, most ($99 \%$) of the gravitational energy released is taken by neutrinos of all flavors in a small burst lasting about ten seconds. As many as 10$^{58}$ neutrinos are emitted rendering the self-interaction sizeable. Similar fluxes are found in simulations of accretion disks around compact binary systems.
Several conversion mechanisms have been uncovered, such as 
the bipolar and the spectral split (see e.g. \cite{Duan:2010bg}), the matter neutrino resonance \cite{Malkus:2012ts} or the "fast" modes \cite{Sawyer:2015dsa}. 
The most general mean-field equations present two more kinds of contributions, i.e. wrong helicity terms due to the neutrino absolute mass \cite{Vlasenko:2013fja} and pairing correlators \cite{Volpe:2013uxl}. They introduce neutrino-antineutrino coupling named helicity coherence that requires anisotropic media to be non-zero. 
Their impact on flavor evolution has been first studied with a one-flavor schematic model in Ref.\cite{Vlasenko:2014bva}. The authors have shown that 
helicity coherence could produce significant flavor modification. An in-depth study based on detailed simulations in binary neutron star mergers has shown that adiabaticity is not enough
to trigger efficient flavor conversion due to helicity coherence. An analytical argument is given to show that this would require a matching between the matter and the self-interaction terms that cannot take place unless specific matter density profiles are chosen. It also shows the requirements for multiple MSW resonances to take place as for the neutrino-matter resonance.

\section{Observational aspects and nuclear physics}
Neutrino observatories often exploit nuclei to detect astrophysical neutrinos.
Scintillator detectors such as JUNO \cite{j} use carbon, water Cherenkov like Super-Kamiokande or the future Hyper-Kamiokande use oxygen, LVD has iron nuclei \cite{Agafonova:2015pca}, DUNE is based on argon \cite{LoSecco:2017bbd} and the HALO observatory uses lead \cite{Vaananen:2011bf}.
The SNEWS network of observatories is ready to observe a future core-collapse supernova explosion \cite{Antonioli:2004zb}.

Neutrino-nucleus cross sections are still affected by significant uncertanties. The best known ones are those on deuterium, used in the SNO experiment, since they can be calculated using Effective Field Theories (see e.g. \cite{Balantekin:2003ep}). The exclusive and inclusive neutrino-carbon cross section associated with neutrinos from decay-at-rest muons have been measured by a series of experiments and are theoretically well understood (see e.g. \cite{Volpe:2000zn}). For the other nuclei, theoretical predictions based on various approaches (mostly shell model, RPA and its variants) have been used over the years\footnote{A measurement on $\nu$-deuterium and one on $\nu$-iron are also available, although the statistical and systematic errors are still large.}.
The operators involved in the nuclear matrix elements of neutrino-nucleus cross sections are of Fermi type, or of Gamow-Teller type. Constraints on these calculations can come from other weak processes such as beta-decay, muon capture, or from charge-exchange reactions. The flux-averaged neutrino-nucleus cross sections associated with typical quasi-thermal supernova neutrino fluxes\footnote{The supernova neutrino spectra are approximated by Fermi-Dirac or power law distributions.} can vary by several tens of percent, e.g. for the case of lead \cite{Vaananen:2011bf} while differences can be even larger for the energy dependent cross sections.

Several proposals have been made over the years 
to measure neutrino-nucleus cross sections of interest for neutrino astrophysics. These include projects such as ORLAND at SNS based on muon decay-at-rest neutrinos, and low-energy
beta-beams that would exploit the beta-decay of boosted radioactive ions to produce neutrinos in the 100 MeV energy range \cite{Volpe:2003fi}.
Experiments to precisely measure neutrino-nucleus cross sections for oxygen, argon, iron and lead are now planned by the COHERENT collaboration at SNS \cite {kate}. Such measurements
will provide information on the isospin and spin-isospin response in these nuclei. In particular they will give further clues on the quenching of the axial-vector coupling constant that is essential for the search of neutrinoless double beta decay \cite{Suhonen:2017krv,Iachello:2015ejm}.
  
The future measurement of supernova neutrinos will bring important information both for astrophysics and neutrino properties. One of the key questions is the mechanism under which iron core-collapse supernovae explode. The favored scenario is the neutrino delayed accretion shock mechanism by Wilson \cite{Wilson:1971} and Bethe-Wilson \cite{Bethe:1984ux}. This is thought to drive most of supernova explosions except probably the most energetic ones. SN1987A observations 
have in particular supported the delayed accretion shock model against the prompt shock one, as shown in the Bayesian analysis of Loredo and Lamb \cite{Loredo:2001rx} and those that have followed \cite{Pagliaroli:2008ur}. 

Neutrinos and antineutrinos of all flavors are emitted during the collapse lasting about ten seconds and take away most of the gravitational binding energy of the newly formed neutron star (about $3 \times 10^{53}$ erg). Current three-dimensional simulations include sophisticated neutrino transport, convection and hydrodynamical instabilities such as the Standing-Accretion-Shock-Instability (SASI). Future observation of supernova neutrinos could confirm/refute the current paradigm where the explosion is aided by the SASI by measuring its imprint on the neutrino time signal in detectors such as ICECUBE (see e.g. \cite{Lund:2010kh}). 

When neutrinos decouple from the dense supernova region and start free streaming, forward scattering that is included in the mean-field approximation produces flavor conversion and swapping(s) among the electron and the muon and tau neutrinos
(see Section 2). These modify the quasi-thermal spectra 
that neutrinos have at decoupling. Reconstructing precisely the neutrino spectra is necessary to get insight on neutrino propagation in the dense region and on the flavor phenomena that occur in the low density layers. Several likelihood analyses have been performed to assess how precisely the neutrino spectra can be reconstructed in the case a supernova explodes. Their determination requires the neutrino luminosity, 
average energy and width (or pinching parameter) for the three flavors (nine free parameters overall) for quasi-thermal distributions. 
In particular, Ref.\cite{Minakata:2008nc} has shown that the detection of neutrinos through inverse-beta decay in Hyper-Kamiokande allows   
to identify the average energies for electron antineutrinos and non-electron neutrino flavors while the presence of the pinching parameters introduces degeneracies that cannot be resolved to fully determine the widths and even the total luminosity. Other analyses have shown that, in principle, all parameters can be precisely measured by combination of several channels such as e.g. in JUNO. However many likelihood analyses make ansatz for example on an energy equipartition
among the neutrino flavors, that is currently not favored by supernova simulations, or by fixing the pinching parameter. Ref.\cite{GalloRosso:2017hbp} has performed likelihood analyses with the 9 parameters unconstrained in order to assess how precisely the emitted gravitational energy of the newly formed neutron star can be determined. These analyses have shown that the combination of inverse-beta decay and electron scattering events allows a measurement of the total gravitational energy with $11 \%$ precision. From such a measurement, and using the relation by Lattimer and Prakash based on EOS that allow for neutron stars with more than 1.65 solar masses, the compactness of the neutron star can be obtained \cite{GalloRosso:2017hbp}. Similar analysis is performed for Hyper-Kamiokande showing that the electron anti-neutrino and non-electron spectra can be precisely determined \cite{GalloRosso:2017mdz}. Hopefully a future measurement of 
neutrinos from a galactic explosion will tell us more about the mass-radius relation and neutron star equation of state.

In conclusion, neutrinos keep revealing intriguing new phenomena in astrophysical environments and interesting connections with other domains, in particular nuclear physics, some of which have been highlighted in this proceedings.




\begin{figure}\label{fig:constraints}
\centering
\includegraphics[width=0.48\textwidth]{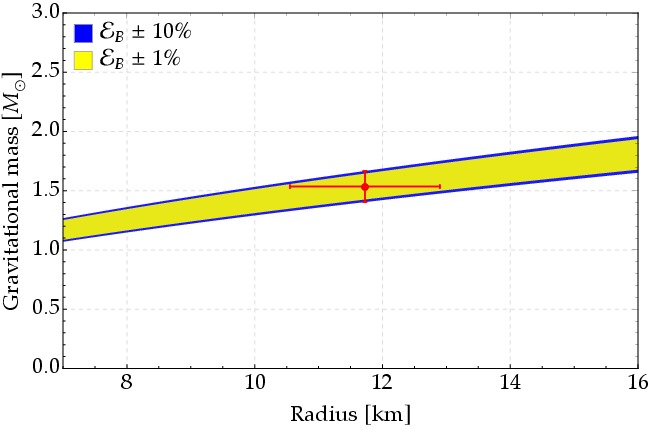}
\caption{Gravitational neutron star mass versus radius allowed region during a core-collapse supernova explosion.
The red point is a putative measurement of the neutron-star radius and the 
corresponding neutron star mass allowed by different EOS following the Lattimer-Prakash relation \cite{Lattimer:2000nx}
(see text). The band is assuming the total  gravitational binding energy of the neutron star emitted with neutrinos is measured with a precision of $10 \%$ with Super-Kamiokande  or $1 \%$ in future detectors like Hyper-Kamiokande. From Ref.\cite{GalloRosso:2017hbp}.}
\end{figure}

\subsection{Acknowledgments}
The author acknowledges financial support from "Gravitation et physique fondamentale" (GPHYS) of the {\it Observatoire de Paris}.

\end{document}